\documentclass[letterpaper, conference]{IEEEtran}

\usepackage[backend=bibtex,
     bibstyle=ieee,
     sortcites=true,
     maxnames=2,
     maxbibnames=12,
     minnames=1,
     mincrossrefs=99,
     uniquelist=false,
     defernumbers=true
]{biblatex}

\AtEveryBibitem{\clearlist{location}}

\makeatletter
\newcounter{IEEE@bibentries}
\renewcommand\IEEEtriggeratref[1]{%
 \renewbibmacro{finentry}{%
   \stepcounter{IEEE@bibentries}%
   \ifthenelse{\equal{\value{IEEE@bibentries}}{#1}}
   {\finentry\@IEEEtriggercmd}
   {\finentry}%
 }%
}
\makeatother

\usepackage{booktabs}
\usepackage{tabularx}
\usepackage{csquotes}

\usepackage{tikz}
\usepackage{tikzpeople}
\usetikzlibrary{trees}
\usetikzlibrary{positioning}
\usetikzlibrary{arrows.meta,arrows}
\usetikzlibrary{decorations.pathreplacing}
\usepackage[edges]{forest}
\forestset{qtree/.style={for tree={parent anchor=south,
           child anchor=north,align=center,inner sep=0pt}}}

\usepackage{pgfplots}
\pgfplotsset{compat=1.18}
\usetikzlibrary{
        pgfplots.dateplot,
        pgfplots.groupplots,
        external,
    }

\usepackage{pgfplotstable}

\usepackage{cleveref}
\crefname{figure}{\figurename}{Tables}
\crefname{table}{\tablename}{Figures}

\usepackage{xspace}
\newcommand*{\eg}{e.g.\@\xspace}
\newcommand*{\ie}{i.e.\@\xspace}

\usepackage{subcaption}

\definecolor{sron0}{HTML}{332288}
\definecolor{sron1}{HTML}{88CCEE}
\definecolor{sron2}{HTML}{117733}
\definecolor{sron3}{HTML}{DDCC77}
\definecolor{sron4}{HTML}{CC6677}
\definecolor{sron5}{HTML}{AA4499}

\begin{document}

\author{\IEEEauthorblockN{Erik Daniel}
\IEEEauthorblockA{
\textit{Technische Universit\"at Berlin}\\
erik.daniel@tu-berlin.de}
\and
\IEEEauthorblockN{Marcel Ebert}
\IEEEauthorblockA{\textit{Technische Universit\"at Berlin}\\
mail@marcel-ebert.de}
\and
\IEEEauthorblockN{Florian Tschorsch}
\IEEEauthorblockA{\textit{Technische Universit\"at Berlin}\\
florian.tschorsch@tu-berlin.de}
}

\title{Improving Bitswap Privacy with \\ Forwarding and Source Obfuscation}

\maketitle

\begin{abstract}
IPFS is a content-addressed decentralized peer-to-peer data network, using the Bitswap protocol for exchanging data.
The data exchange leaks the information to all neighbors, compromising a user's privacy.
This paper investigates the suitability of forwarding with source obfuscation techniques for improving the privacy of the Bitswap protocol.
The usage of forwarding can add plausible deniability and the source obfuscation provides additional protection against passive observers.
First results showed that through trickle-spreading the source prediction could decrease to 40\,\%, at the cost of an increased content fetching time.
However, assuming short distances between content provider and consumer the content fetching time can be faster even with the additional source obfuscation.
\end{abstract}


\section{Introduction}
The InterPlanetary File System~(IPFS)~\cite{benet2014ipfs} is a peer-to-peer~(P2P) overlay network for distributed storage and exchange of files.
Data is content-addressed and exchanged using the Bitswap protocol.
Bitswap asks immediate neighbors for possession of the desired content.
Additionally, content can be located with the help of a Kademlia-based DHT, in case no neighbor is in possession of the content.
This procedure allows a passive observer to identify and track a peer's interest~\cite{balduf2021monitoring}.

In this paper, we investigate whether request spreading, \ie, trickle spreading, can help to increase the privacy of a peer.
The addition of request forwarding introduces plausible deniability.
The source obfuscation should protect the plausible deniability against a passive observer, improving privacy.
Trickle-spreading or trickling is a gossip-based flooding protocol~\cite{fanti2017anonymity} that aims to protect the source of a request from a passive observer.
While trickling has shown limited benefits~\cite{neudecker2016timing,fanti2017anonymity} in the Bitcoin network, it improves protection.

In order to investigate the feasibility of trickling,
we integrated request forwarding with trickling into \texttt{go-bitswap} providing a Proof-of-Concept~(PoC) implementation.
The PoC is evaluated using different simulations.
The simulations show the privacy gains and the performance trade-off for different network and trickling parameters.
The results confirm previous simulations~\cite{de2021accelerating}
showing that forwarding can improve content retrieval time.
Even with trickling, the retrieval time can be lower than the default behavior. 
Furthermore, trickling can reduce the source prediction accuracy of a single passive observer to $40\,\%$,
although, increasing retrieval time.

\section{Source obfuscation}
The added privacy consists of two components: forwarding and source obfuscation. 
By integrating forwarding into the data request, every peer gains plausible deniability for requests,
because every request could potentially be made on behalf of other peers.
However, a passive observer can identify the source of the request by listening to many peers at the same time~\cite{balduf2021monitoring}.
Therefore, forwarding alone is not sufficient.
Source obfuscation can add protection against these passive observers.
This is a common problem in cryptocurrencies,
where it is desired to protect the origin of transactions.
Cryptocurrencies use among others spreading protocols like trickling.

Trickling is a gossip-based flooding protocol~\cite{fanti2017anonymity} and used to broadcast messages.
The trickling can be interpreted as relaying a message in rounds.
Each peer orders its neighbors, which are assumed to not know the message.
In the first round, the peer selects a group of peers based on the order, which receive the message immediately.
Each following round starts after a fixed delay in which the next group of peers receive the message.
The size of the group could be adjusted according to the size of the network and the number of neighbors.

In the following, we describe the Bitswap protocol and the changes to the default behavior due to forwarding and trickling.
For the evaluation, the described changes are implemented as a PoC.
The PoC implementation is based on \texttt{go-bitswap}~\cite{gitgobs} v0.10.0, and publicly available.%
\footnote{https://github.com/ebma/go-bitswap}

\subsection{Forwarding in Bitswap}
In IPFS, Bitswap is responsible for the exchange of data.
The smallest data item is a block, which is identified with a content identifier, a CID.
Bitswap sends as a first step, to discover possible content providers, to each neighbor a \texttt{WANT-HAVE} request for the CID.
The \texttt{WANT-HAVE} is answered with a \texttt{HAVE} in case a peer has the block and can be answered with a \texttt{DONT-HAVE} in case the peer does not have the block.
After receiving a \texttt{HAVE}, a peer can request the block with \texttt{WANT-BLOCK} and receives a \texttt{BLOCK} with the data.
In case a peer is no longer interested or received the block from another peer, the peer sends a \texttt{CANCEL}.

In the Bitswap context, forwarding mainly means to forward the \texttt{WANT-HAVE}s.
This requires to change the handling in case the peer does not have the requested block.
That is, a relaying peer handles remote \texttt{WANT-HAVE} request as if they are its own requests, relaying the request and fetching the block in case a content provider is found.
The return of the block to the requesting source is already part of the protocol.
When a peer receives a block, it announces possession of the block to other interested peers, notifying them.
Trickling changes sending of \texttt{WANT-HAVE} requests and \texttt{CANCEL} notifications.

To this end, some modifications of the Bitswap code are required.
Forwarding was previously investigated by \citeauthor{de2020teaching}~\cite{de2020teaching}, which is taken as the basis for this PoC.
In Bitswap, sessions track peer requests and coordinate messages exchanged with other peers.
Therefore, a relay session is added to Bitswap, the relay session tracks all \texttt{WANT} requests and CIDs that are requested on behalf of other peers~\cite{de2020teaching}.
A peer receiving a \texttt{WANT} request for an unknown CID responds with a \texttt{DONT-HAVE} and adds the interest of the peer in the relay session.
When the relay session is updated, a \texttt{WANT-HAVE} request is sent to the other peers, forwarding the message.
Peers receiving a new block check the relay session if a peer is interested in the block, forwarding the block to interested peers.
Either receiving a block or a \texttt{CANCEL} removes the peer's interest in the CID from the relay session.
Contrary to normal sessions, relay sessions do not resend \texttt{WANT-HAVE} requests or search for new providers.
The relay sessions are identified by a boolean parameter added to the normal sessions.
To simplify the forwarding logic, Bitswap adds blocks to the client's blockstore, instead of the caller of the service.
Additionally, the session manager is made to be shared by the client and server component of Bitswap, since the decision engine decides the creation and update of relay sessions.
The session manager periodically checks sessions interests in blocks, sending \texttt{CANCEL}s for canceled or fulfilled interests.
Therefore, \texttt{CANCEL} requests are automatically forwarded.

\subsection{Trickling in Bitswap}
Trickling changes the sending of requests to a peer's neighbors.
The peer-want manager, a subcomponent of the peer manager, keeps track of \texttt{WANT-HAVE} and \texttt{WANT-BLOCK} requests sent to peers.
It is also responsible for sending \texttt{CANCEL} request.
During the default operation of Bitswap, a peer wants to request the CID from all neighbors.
Therefore, the peer-want manager checks whether the request was transmitted before, queuing the request into the message queues of all non-transmitted peers.
This queuing needs to be trickled.

A mutual exclusion lock of the peer-want manager is locked before each \texttt{WANT} request is sent.
The lock is unlocked after pausing the Go-routine for the duration of the pre-defined trickling delay.
The order of the peers is randomized for every iteration with a permutation from a pseudo-random number generator.
The seed for the pseudo-random number generator is the timestamp on initialization.
This randomization allows to change the order in which peers receive the request.
The same is done for the transmission of \texttt{CANCEL} requests.

Originally, \citeauthor{de2020teaching}~\cite{de2020teaching} added a TTL to every Bitswap message.
The TTL is reduced by 1 before forwarding the message until the TTL reaches 0.
This reduces the load of the network due to multiple redundant sent and received messages.
A fixed TTL can reveal the source of the request, making the addition of trickling useless.
To keep the disappearing effect of the TTL, it is possible to alter the change and initial value of the TTL.
A peer could randomly change the TTL, although the TTL should ultimately decline.
Another possibility is to change the forwarding independent of a TTL, \eg each peer forwards a message only with a certain probability.
The PoC implementation does not contain a TTL.
The missing TTL, trickling and adoption to a newer Bitswap version are the main differences to~\cite{de2020teaching}.

\section{Simulation Environment}
The impact of trickling on the performance and the privacy gains are evaluated with simulations, using Testground.\footnote{\url{https://docs.testground.ai/} (Accessed: 2023-03)}
Testground is an open source tool for testing P2P systems.
Simulations are executed by testplans.
We used two different testplans: one for evaluating the trickle-forwarding modifications
and one for go-bitswap with IPFS.
The trickle-forwarding is used to analyze the privacy and performance,
while Bitswap with IPFS serves as a performance baseline.
The testplans are based on the Beyond Bitswap testbed.\footnote{\url{https://github.com/protocol/beyond-bitswap} (Accessed:2023-03)}

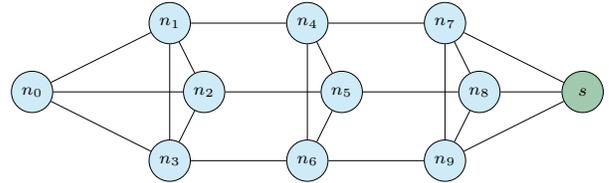
\begin{figure}
    \footnotesize\centering
    \resizebox{0.9\columnwidth}{!}{
    \begin{tikzpicture}
        \tikzstyle{leech}=[circle, draw, thin,fill=sron0!40, minimum width=6mm, font=\scriptsize]
        \tikzstyle{passive}=[circle, draw, thin,fill=sron1!40, minimum width=6mm, font=\scriptsize]
        \tikzstyle{seed}=[circle, draw, thin,fill=sron2!40, minimum width=6mm, font=\scriptsize]

        \node [passive] (a) at (0, 0) {$n_0$};
        \node [passive] (b) at (2, 1) {$n_1$};
        \node [passive] (c) at (2.5, 0) {$n_2$};
        \node [passive] (d) at (2, -1) {$n_3$};
        \node [passive] (e) at (4, 1) {$n_4$};
        \node [passive] (f) at (4.5, 0) {$n_5$};
        \node [passive] (g) at (4, -1) {$n_6$};
        \node [passive] (h) at (6, 1) {$n_7$};
        \node [passive] (i) at (6.5, 0) {$n_8$};
        \node [passive] (j) at (6, -1) {$n_9$};
        \node [seed] (k) at (8, 0) {$s$};

        \path[thin] (a) edge (b);
        \path (a) edge (c);
        \path[thin] (a) edge (d);
        \path[thin] (b) edge (c);
        \path[thin] (b) edge (d);
        \path[thin] (b) edge (e);
        \path[thin] (c) edge (d);
        \path[thin] (c) edge (f);
        \path[thin] (d) edge (g);
        \path[thin] (e) edge (f);
        \path[thin] (e) edge (g);
        \path[thin] (e) edge (h);
        \path[thin] (f) edge (g);
        \path[thin] (f) edge (i);
        \path[thin] (g) edge (j);
        \path[thin] (h) edge (i);
        \path[thin] (h) edge (j);
        \path[thin] (h) edge (k);
        \path[thin] (i) edge (j);
        \path[thin] (i) edge (k);
        \path[thin] (j) edge (k);
    \end{tikzpicture}}
    \caption{Simulation network topology without eavesdropper.}
    \label{fig:topo}
\end{figure}

For comparison, one network topology is used for all tests,
which is illustrated in \cref{fig:topo}.
The network consists of 11 honest peers in total
and are divided in one seed (providing data),
one leech (requesting data), and 9 passive forwarding nodes.
We use two scenarios that differ only in the position of the leech:
the center scenario (node~$n_5$ is leech),
and the edge scenario (node~$n_0$ is leech).
In addition, a varying number of eavesdropper nodes
(\ie, colluding, passive, non-forwarding observer) are added,
each directly connected to \emph{all} nodes.
While desirable, more complex scenarios with more nodes was not feasible
with the given hardware
(Intel Core i7-12700K CPU with 32\,GB RAM,
Samsung 980 PRO SSD,
and 64\,b Manjaro with Linux Kernel 5.15.65-1).

In the trickle-forwarding testplan, each node runs a libp2p host with a modified Bitswap.
Each node has a TCP and a QUIC multiaddress.
Since QUIC has a higher priority, TCP is only used as a fail-safe.
The seed node publishes a predefined test file and communicates the root CID to the leech.
The test file is a randomly created file of a predetermined size.
The leech fetches the test file.
Between each run, all nodes are re-instantiated,
clearing the blockstore and pending messages from preceding runs.
For each run, the time to fetch the file and relevant metadata is recorded.
The number of runs and eavesdropper, link latency, and trickling delay can be set.

The testplan with the unmodified go-bitswap and IPFS
does not use forwarding or trickling.
Instead of a libp2p node with a modified Bitswap,
each node deploys a kubo\footnote{https://github.com/ipfs/kubo} node, an IPFS implementation in Go.
Kubo nodes are used for a simpler fallback of using the DHT for finding the content provider.
Since this testplan is only used for performance comparisons, no eavesdropper and no trickling is configured.
The number of runs and the link latency can be configured.

\section{Evaluation}
We simulated each configuration
with a link latency of $50\,ms$, $100\,ms$, and $150\,ms$
and trickling delays in the range of $0\,ms$--$300\,ms$ in $50\,ms$ steps.
Furthermore, three different test file sizes were analyzed:
$512\,B$, $150\,KiB$, and $1\,MiB$.
The small file was chosen,
since \texttt{WANT-HAVE} requests for blocks smaller than $1\,KiB$
are directly answered with a block.

\subsection{Privacy}
Trickling and forwarding should provide plausible deniability for a client as well as source obfuscation from a passive observer.
To analyze the effectiveness, the adversarial model of eavesdroppers is used as introduced by~\citeauthor{fanti2017anonymity}~\cite{fanti2017anonymity} for analyzing the Bitcoin network.
In the model, a supernode connects to most, ideally all, peers in the network, possibly establishing multiple connection to each peer.
The supernode \enquote{eavesdrops} on all messages and logs the timestamps of received messages without further processing.
In the simulation multiple connections are replaced with multiple nodes each establishing one connection, the eavesdroppers.

A first-timestamp estimator compares the timestamp and outputs the first-timestamp as the source.
This can already achieve high accuracy rates as shown by~\cite{biryukov2014deanonymisation,koshy2014analysis}.

\begin{figure}
    \includegraphics[width=0.5\textwidth]{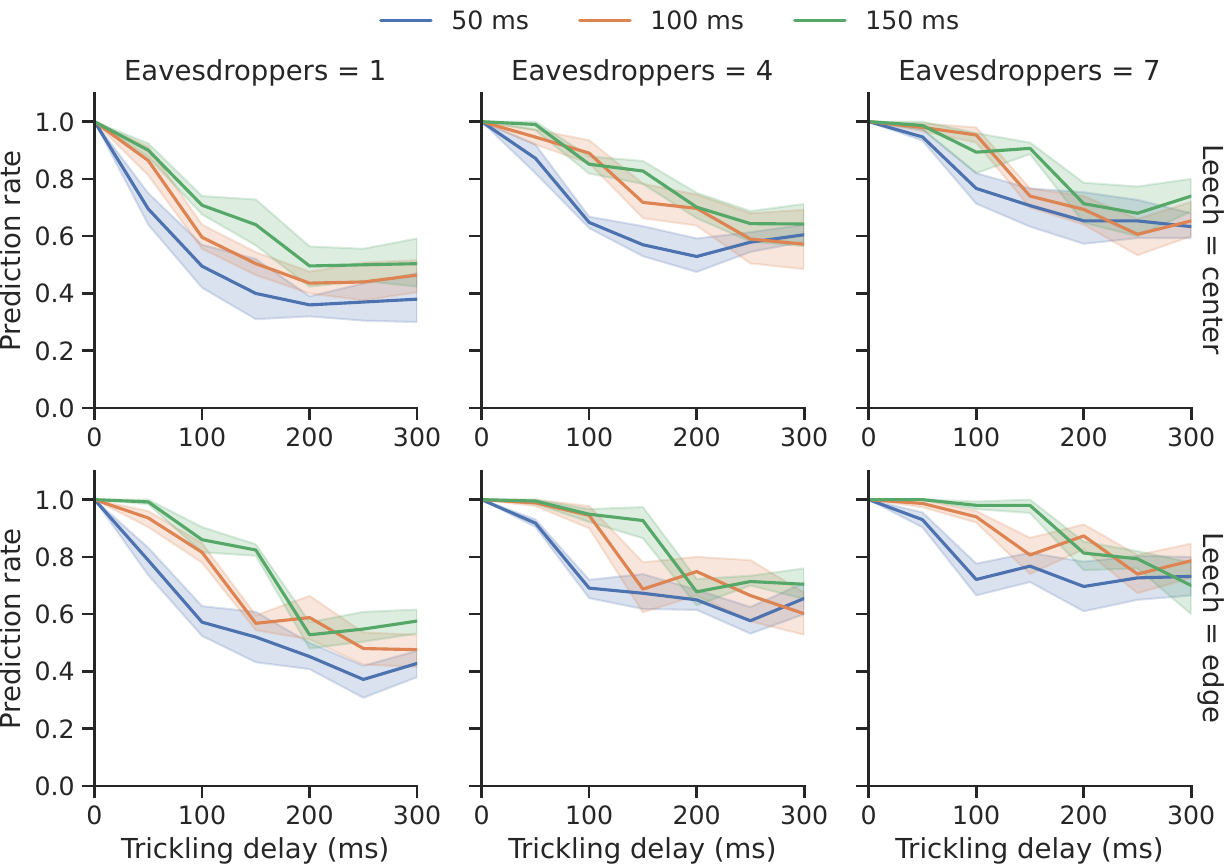}
    \caption{Prediction Accuracy.}
    \label{fig:pa}
\end{figure}

For the results, the prediction accuracy is calculated by dividing the correct predictions of the estimator by the total number of runs for the selected parameters.
Each parameter configuration is executed at least 3 times for both network configurations with 50 runs for $1$ eavesdropper, $40$ runs for $4$ eavesdropper, and $30$ runs for $7$ eavesdropper.
The results are the average prediction accuracy and can be seen in \cref{fig:pa}.

It can be seen that the prediction rate declines with a higher trickling delay.
The prediction accuracy drops to $40\,\%$ for a single eavesdropper, while the accuracy remains between $60\,\%$ and $80\,\%$ for $7$ eavesdropper.
The prediction accuracy seems to correlate to the latency, \eg, $80\,\%$ for trickling delay of $50\,ms$ with a $50\,ms$ latency as well as a trickling delay of $150\,ms$ with a $150\,ms$ latency.
As expected, the prediction accuracy increases with a higher number of eavesdropper nodes.
However, the prediction accuracy is higher in case of the edge network.
This is due to the longer path and slower message propagation, giving the adversary more chances to receive the message from the source first.
It seems that the adversary has a higher chance to identify the source in case of higher network latency, however, this could be due to the correlation of trickling delay and latency.

\subsection{Performance}
\begin{figure}
    \includegraphics[width=0.5\textwidth]{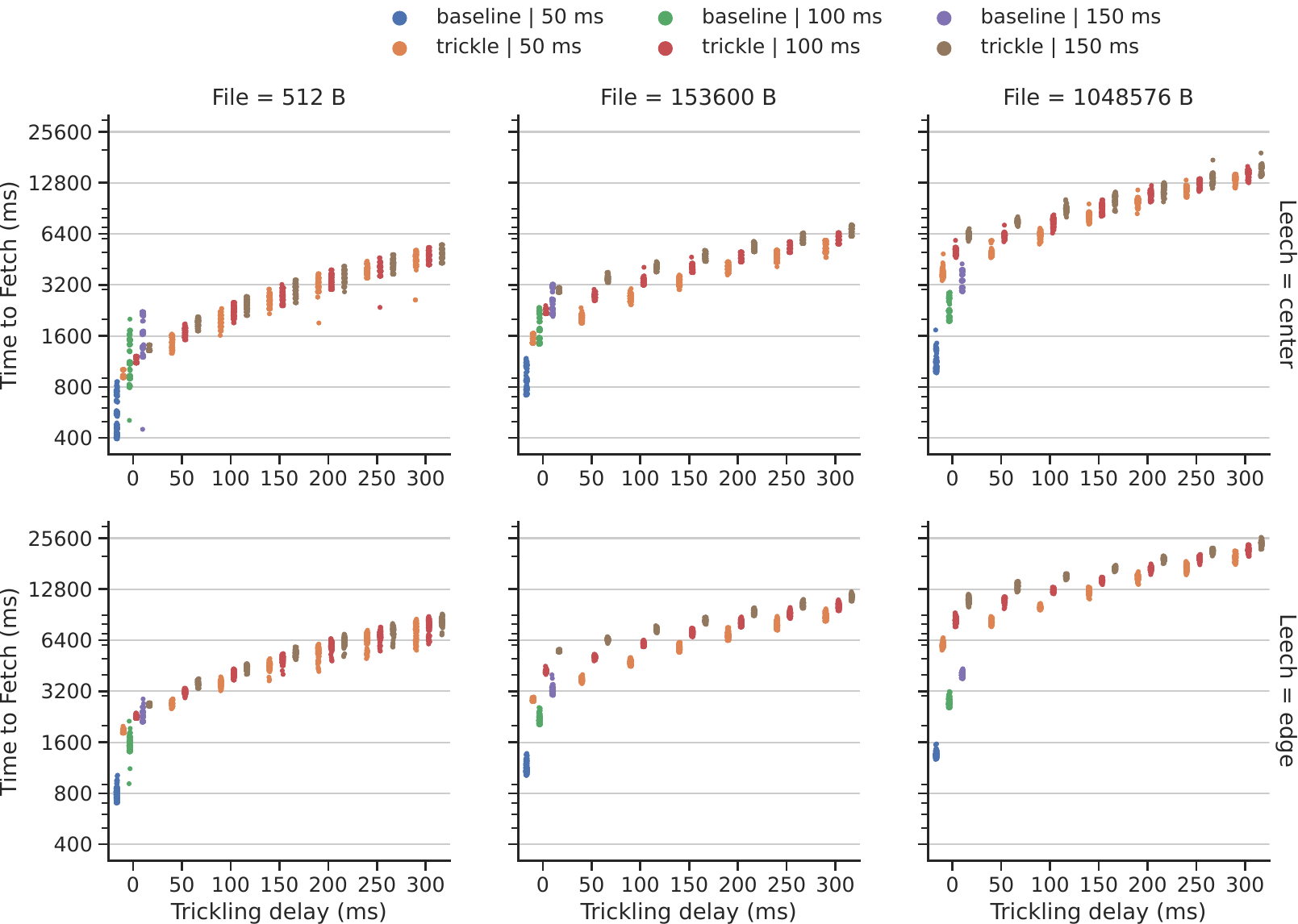}
    \caption{Time to fetch.}
    \label{fig:perf}
\end{figure}

For the performance evaluation, each simulation is run multiple times with multiple runs for the baseline and the different file sizes.
The results can be seen in \cref{fig:perf}.
The time to fetch~(TTF) with trickling always takes more time than the baseline experiments, except for short distances, \eg, with the leech in the center and low trickling delays.
This time advantage can only be gained with smaller files resulting in only a few block fetches.
For more blocks the normal variant has the advantage of being directly connected to the provider not requiring any additional DHT lookups.
While forwarding even with trickling can speed up content fetching, depending on the distance, it can considerably slow down content fetching time.
Additionally, due to the default minimum number of 10 peers in the routing table and the small number of nodes in the experiment, the address of the seed is most likely known by all nodes further speeding up the DHT lookup.

\subsection{Discussion}
In summary, the ideal result for the privacy evaluation would be that every node has the same probability to be the source of a request, \ie, 1/N.
For the simulation that would mean a prediction accuracy of $\approx 10\,\%$.
The results show values far above the ideal result.
Yet, there is some improvement to a client's privacy
compared to the default behavior,
where the eavesdropper can determine the source of any request with certainty.
It should be noted, that if trickling is used, it should be adjusted to the network latency to improve its effectiveness.
Considering the performance, the simulation can confirm the results of other work~\cite{de2021accelerating} that forwarding has the potential to speed up content retrieval.
However, the forwarding range should be limited to prevent flooding the network, especially since the speedup of forwarding seems to decrease with the distance of requestor and source, and the number of blocks.

While the results seem promising, the fixed topology with limited number of nodes and defined link latency, is not directly comparable to the IPFS network~\cite{henningsen2020mapping,daniel2022passively,trautwein2022design}.
There are much more nodes with more neighbors.
Therefore, an adoption to the IPFS network would require more tests concerning the trickling delay, forwarding reach, and number of contacted peers per trickling round.
However, the results provide a first estimation to the feasibility of trickling.

\section{Related Work}
Forwarding for Bitswap was already investigated by \citeauthor{de2021accelerating}~\cite{de2021accelerating} with the goal of improving retrieval times.
This work investigated the addition of trickling and forwarding to improve the privacy.
There are other source obfuscation techniques, which could also be used as alternatives.
In diffusion spreading, the delay is independently drawn from an exponential distribution.
Since the prediction accuracy of trickling showed a dependency of trickling delay and latency,
a random delay might be more effective in a real network.
Although, simulations for the Bitcoin network showed similar detection properties for diffusion spreading and trickling~\cite{fanti2017anonymity}.
Another approach to obfuscate the source is a random walk as proposed in Dandelion~\cite{bojja2017dandelion} and Dandelion++~\cite{fanti2018dandelion++}.
However, the provided anonymity of Dandelion/Dandelion++'s transaction broadcast mechanism is also limited~\cite{sharma2022anonymity}.

Forwarding and source obfuscation can improve the privacy of content retrieval by adding a considerable load to the network.
As an alternative request forwarding could be very limited and only used for proxy selection.
The request is forwarded a few hops after which a peer requests the content on behalf of the original node, acting as a proxy.
After the proxy retrieved the content, it is routed back to the original node.
This proxy selection can be done, \eg, with a random walk as proposed by~\cite{qian2016garlic}.

\section{Conclusion}\label{sec:conclusion}
In this paper, we showed a possibility to improve
content retrieval privacy by utilizing request forwarding and trickling.
Forwarding requests itself can speed up content retrieval, improve content availability and resilience, and introduces some plausible deniability.
Trickling at the cost of possible performance gains can further improve the privacy compared to pure forwarding.
However, the effectiveness of trickling is limited and needs fine-tuning of many parameters, and the presented evaluation provides only a limited impression of the effectiveness of trickling.
Furthermore, forwarding should be limited to some degree to reduce the strain on the network.

\section*{Acknowledgements}
We thank Protocol Labs for funding our research.

\printbibliography[heading=bibintoc]

\end{document}